\documentstyle[12pt,amsfonts]{article}
\def\be{\begin{equation}}
\def\ee{\end{equation}}
\def\ba{\begin{array}}
\def\ea{\end{array}}
\begin{document}
\title{\bf Problems on Foundations of General Relativity }
\author{{Ning Wu$^1$\thanks{email address: wuning@mail.ihep.ac.cn} ,
Tu-Nan Ruan$^{2,3}$}  \\
{\small Institute of High Energy Physics, P.O.Box 918-1,
Beijing 100039, P.R.China$^1$} \\
{\small CCAST(World Lab), Beijing  100080, P.R.China $^2$}\\
{\small Dept. of Modern Phys., Univ. of Sci. and
Tech. of China, Hefei, Anhui 230026, P.R.China $^3$  }  }
\maketitle
\vskip 0.8in

~~

\vskip 0.8in

\begin{abstract}
It was generally believed that, in general relativity, the fundamental
laws of nature should be invariant or covariant under a general
coordinate transformation. In general relativity, the equivalence
principle tells us the existence of a local inertial coordinate
system and the fundamental laws in the local inertial coordinate system
which are the same as those in inertial reference system. Then, after a
general coordinate transformation, the fundamental laws of nature
in arbitrary coordinate system or in arbitrary curved space-time can
be obtained.  However, through a simple example, we find
that, under a general coordinate transformation, basic physical
equations in general relativity do not transform covariantly,
especially they do not preserve their forms under the transformation
from a local inertial coordinate system to a curved space-time.
The origination of the violation of the general covariance is then
studied, and a general theory on general coordinate
transformations is developed.  Because of the the existence of the
non-homogeneous term, the fundamental laws of nature in
arbitrary curved space-time can not be expressed  by space-time
metric, physical observable and their derivatives. In other words,
basic physical equations obtained from the equivalence principle and
the principle of general covariance are different from those
in general relativity.
Both the equivalence principle and the principle of general covariance
can not be treated as foundations of general
relativity. So, what are the foundations of General Relativity?
Such kind of essential problems on General Relativity
can be avoided in the physical picture of gravity.
Quantum gauge theory of gravity, which is founded in the physics
picture of gravity,  does not have such kind of fundamental
problems.
\\

\end{abstract}

\Roman{section}

\section{Introduction}
\setcounter{equation}{0}

In classical theory of gravity, gravity is treated
as a kind of physical interactions
in flat space-time and obeys inverse square law\cite{01}.
In general theory of relativity, gravity is treated as
space-time geometry\cite{02,03}. In order to consider
quantum effects of gravitational interactions, various
kinds of quantum gravity are studies. In some quantum
theories of gravity,
gravity is treated as a kind of interactions in space-time.
Therefore, there are two different ways to treat gravity,
or there are two essentially different points of view on
the nature of gravity: one considers gravity as a kind of
fundamental physical interactions, and another considers
gravity as space-time geometry. For the sake of convenience,
we call the former the physics picture of gravity and the
latter the geometry picture of gravity. \\

For some fundamental problems, two pictures of gravity are
not consistent. For example, in geometry picture of gravity,
there are no physical gravitational interactions in space-time
and all gravity effects are only effects of curved space-time,
while in physics picture of gravity, gravity is
treated as a kind of fundamental physical
interactions in space-time and space-time is always kept flat.
What is the nature of gravity? In other words, gravity
is a kind of fundamental interactions or space-time
geometry? Is our space-time essentially flat or curved?\\

A fundamental theory to describe various kinds of
fundamental interactions in nature is gauge theory\cite{1}.
Electromagnetic interactions, weak interactions and
strong interactions can all be described by gauge
theory. Various kinds of unified theories of fundamental
interactions are also based on gauge theory, such as
unified electroweak theory\cite{2,3,4,401} is a
$SU(2)_L \times U(1)_Y$ gauge theory.
Now it is generally  believed that four kinds of
fundamental interactions in nature are all gauge
interactions and they can be described by gauge field theory.
From theoretical point of view, the principle of local
gauge invariance plays a fundamental role in particle's
interaction theory.  \\

Gauge theory is so successful in describing fundamental
interactions of nature and in unifying different kinds
of fundamental interactions, it is natural for physicist
to study gravity and to unify gravity with other kinds
of fundamental interactions by  gauge field theory.
In 1918, H. Weyl first use gauge theory to study
graivty\cite{41}. Later, gauge gravity is studied
extensively\cite{b1,b2,b3,b31,b34,b4,b5,b6}.
In the traditional
gauge treatment of gravity, Lorentz group is localized, and the
gravitational fields are not represented by gauge
potentials. \\

Recently, a new quantum gauge theory of gravity, which is
perturbatively renormalizable in 4-dimentional space-time,
is proposed by N. Wu\cite{wu1,wu2,wu3,wu4}. In quantum gauge
theory of  gravity, gravity is treated as a kind of fundamental
interactions, so it is formulated in physics picture of
gravity. Quantum gauge theory
of gravity is based on gauge principle,
which is the common nature of all kinds of fundamental
interactions in nature. One of the most important advantage
of quantum gauge theory of gravity is that four different
kinds of fundamental interactions in nature can be unified
in a simple and beautiful way\cite{wu5,wu6,wu7}. Quantum
gauge theory of gravity can be used to explain both quantum
phenomena and classical phenomena of gravitational interactions.
\\

However, basic dynamics of gravitational interactions
given by different theories of gravity is different. There are two ways
to test different kinds of gravity theories. One is to  test
by experiments. Another way is to check self-consistency
of the theory itself. In this paper, the second way is used to
test different kinds of gravity theories.  \\

Another fundamental important problem on gravity theory is that
what is the symmetry of gravitational interactions.
In general relativity, the symmetry of gravitational interactions
is general covariance under a general coordinate transformation.
In most traditional quantum gravity, the symmetry of gravitational
interactions is local Lorentz symmetry. In quantum gauge theory
of gravity, the fundamental symmetry is gravitational gauge
symmetry. So, different theories have different symmetries. However,
the objective gravitational interactions in nature must have
definite symmetry. So, what is the symmetry of gravity? The
theory with incorrect symmetry can not be an acceptable fundamental
theory of gravity. To determine the true symmetry of gravity
will help us to determine which is the fundamental
theory of gravity. \\

In this paper, our main spirit is to study the symmetry of gravity.
Using a simple example, it can be proved that almost all basic physical
equations in general relativity do not transform covariantly
under a general coordinate transformation. In this simple model,
the transformation rule of basic physical equations
can be calculated explicitly. It is found that the transformation
rules given by direct calculation are different from those in
general relativity. So, what is the origin of such kind
of violation? This question is answered by a general theory, which is
developed to study the origin of such kind of violation of general
covariance under a general coordinate transformation. The violation
of general covariance means that both the principle of general
covariance and the equivalence principle, which are believed to be
{\it a priori} foundations of general relativity, are no longer the
foundations of general relativity.
Combining discussions in this paper and results from gauge
principle\cite{wu1,wu2,wu3,wu4}, it will be argued that the
correct symmetry for gravitational interactions is gravitational
gauge symmetry. \\

\section{Foundations of General Relativity}
\setcounter{equation}{0}

Before we study problems on foundations
of general relativity, let's remember
what are the foundations of general relativity. It is well known
that the foundations of general relativity are two principles:
the equivalence principle and the principle of general
covariance\cite{02,03,wei}.  Based on these two principles,
basic physical equations in arbitrary
gravitational field can be obtained.\\

Equivalence principle states that at every space-time point in an arbitrary
gravitational field it is possible to choose a local inertial
coordinate system such that, within a sufficiently small region
of the point in question, the laws of nature take the same form
as in unaccelerated Cartesian coordinate systems in the
absence of gravitation. There are two key points in the physics
content of the equivalence principle: the existence of a local
inertial coordinate system and the form of the laws of nature
in the local inertial coordinate system.
According to the equivalence principle, there must exist a local
inertial coordinate system at every space-time point in an
arbitrary curved space-time, and the forms of the laws of nature
in the local inertial coordinate system are the same as those
in inertial reference system. It is known that the laws of nature
in a inertial reference system are given by special relativity.
According to special relativity, basic physical equations
in a inertial reference system are
\be  \label{2.1}
\partial_{\mu} J^{\mu} = 0,
\ee
\be  \label{2.2}
\frac{{\rm d}^2 \xi^{\mu}}{{\rm d} \tau ^2} = 0,
\ee
\be  \label{2.3}
\frac{{\rm d} A^{\mu}}{{\rm d} \tau } = 0,
\ee
\be  \label{2.4}
\partial_{\nu} T^{\mu \nu} = 0,
\ee
where $J^{\mu}$ is a conserved current,
$\xi^{\mu}$ is the space-time coordinate of a mass point,
$A^{\mu}$ is an arbitrary vector,
$T^{\mu \nu}$ is
energy-momentum tensor and $\tau$ is the proper time.
Eq.(\ref{2.1}) is the continuity equation, eq.(\ref{2.2})
is the equation of motion for a free mass point,
eq.(\ref{2.3}) is the parallel transport equation and
eq.(\ref{2.4}) is the energy-momentum conservation
equation. Equivalence principle tells us that these
equations hold in a local inertial coordinate system.\\

But the equivalence principle does not directly tell us what are
the forms of the laws of nature in an arbitrary curved space-time.
This task is accomplished by the principle of general covariance.
The principle of general covariance states that a physical
equation holds in a general gravitational field, if two
conditions are met: (1) The equation holds in the absence of
gravitation; that is, it agrees with the laws of special relativity
when the metric tensor $g_{\alpha \beta}$ equals the Minkowski
tensor $\eta_{\alpha \beta}$ and when the affine connections
$\Gamma^{\alpha}_{\beta \gamma}$ vanishes. (2) The equation
is generally covariant; that is, it preserves its form under
a general coordinate transformation. According to
the principle of general covariance, the basic physical
equations in an arbitrary curved space-time are
\be  \label{2.5}
\nabla_{\alpha} J^{\alpha} = 0,
\ee
\be  \label{2.6}
\frac{{\rm d}^2 x^{\alpha}}{{\rm d} \tau ^2}
+ \Gamma^{\alpha}_{\beta \gamma}
\frac{{\rm d} x^{\beta}}{{\rm d} \tau}
\frac{{\rm d} x^{\gamma}}{{\rm d} \tau} = 0,
\ee
\be  \label{2.7}
\frac{{\rm D} A^{\alpha}}{{\rm D} \tau } = 0,
\ee
\be  \label{2.8}
\nabla_{\beta} T^{\alpha \beta} = 0,
\ee
where $\nabla_{\alpha}$ is a covariant derivative,
$\frac{{\rm D}}{{\rm D} \tau}$ is a covariant derivative
along the curve $x^{\mu}(\tau)$ which is defined by
\be  \label{2.9}
\frac{{\rm D} A^{\alpha}}{{\rm D} \tau }
= \frac{{\rm d} A^{\alpha}}{{\rm d} \tau }
+  \Gamma^{\alpha}_{\beta\gamma}
\frac{{\rm d}x^{\gamma}}{{\rm d} \tau} A^{\beta}.
\ee
It is generally believed that eqs.(\ref{2.5}) -- (\ref{2.8})
transform covariantly under a general coordinate transformation,
so they preserve their forms under a general coordinate
transformation; that is, these equations hold in an arbitrary
curved space-time. \\

The key point in the principle of general covariance is that
all basic physical equations should be covariant under a
general coordinate transformation. It is generally believed
that all physical equations, including Einstein field equation,
are covariant under a general coordinate transformation, so they
have the same form in all possible curved space-times and all
possible coordinate systems. If a physical equation does not
transform covariantly, it will have different forms in
different coordinate systems. In this case, all
basic  physical equations
are space-time dependent, and we can not simple write out an
equation which will hold in an arbitrary coordinate system.  \\

Though it was generally believed that Einstein field equation
\be  \label{2.10}
R_{\alpha \beta} -  \frac{1}{2} g_{\alpha \beta} R
= - 8 \pi G_N T_{\alpha \beta}
\ee
and eqs.(\ref{2.5}) -- (\ref{2.6}) are covariant under a
general coordinate transformation, through a simple example,
we will show that they are not really  covariant under a
general coordinate transformation, which will cause
serious problems to the foundations of general relativity. \\

\section{A Simple Example}
\setcounter{equation}{0}

For the sake of convenience, coordinates of flat space-time
are denoted by $\xi^{\mu}=(t,r,\theta,\varphi)$
and coordinates of curved space-time
are denoted by $x^{\alpha}=(t', r', \theta ', \varphi ')$.
Greek indices $\mu$, $\nu$, $\kappa$,
$\lambda$, and so on generally run over the four space-time
inertial coordinate labels 0, 1, 2, 3 or $t$, $r$, $\theta$, $\varphi$;
Greek indices $\alpha$, $\beta$, $\gamma$, $\delta$ and
so on generally run over the four coordinate labels in a
general coordinate system. The metric $\eta_{\mu \nu}$
of flat space-time
has diagonal elements -1, +1, $r^2$, $r^2 {\rm sin}^2 \theta$.
All calculations in this chapter are performed in spherical
coordinate system.
\\

Let's discuss a simple coordinates transformation from flat
space-time $\xi^{\mu}$ to curved space-time $x^{\alpha}$:
\be \label{3.2}
{\rm d} \xi^{\mu} \to {\rm d} x^{\alpha}
= \frac{\partial x^{\alpha}}{\partial \xi^{\mu}}
{\rm d} \xi^{\mu},
\ee

\be \label{3.3}
\frac{\partial}{ \partial \xi^{\mu}}  \to
\frac{\partial}{\partial x^{\alpha}} =
\frac{\partial \xi^{\mu}}{\partial x^{\alpha}}
\frac{\partial}{\partial \xi^{\mu}},
\ee

where transformation matrices
$\frac{\partial x^{\alpha}}{\partial \xi^{\mu}}$
and
$\frac{\partial \xi^{\mu}}{\partial x^{\alpha}}$
are defined by
\be \label{3.4}
\frac{\partial \xi^{\mu}}{\partial x^{\alpha}}
= \left (
\ba{cccc}
\sqrt{1-u^2} &0&0& 0 \\
0& \frac{1}{\sqrt{1-u^2}} &0&0  \\
0&0&  1&0\\
0&0&0&  1
\ea
\right ),
\ee

\be \label{3.5}
\frac{\partial x^{\alpha}}{\partial \xi^{\mu}}
= \left (
\ba{cccc}
\frac{1}{\sqrt{1-u^2}} &0&0& 0 \\
0& \sqrt{1-u^2} &0&0  \\
0&0&  1&0\\
0&0&0&  1
\ea
\right ).
\ee
In order to simplify our discussion, in this paper, we only
suppose that $u$ is a arbitrary function of radius $r$;
that is,
\be \label{3.6}
u = u(r),
\ee
and
\be \label{3.7}
\frac{\partial u}{\partial t}
= \frac{\partial u}{\partial \theta}
= \frac{\partial u}{\partial \varphi}
=0.
\ee
\\

It can be easily proved that the transformation matrices
satisfy
\be \label{3.8}
\frac{\partial \xi^{\mu}}{\partial x^{\alpha}}
\frac{\partial x^{\alpha}}{\partial \xi^{\nu}}
= \delta^{\mu}_{\nu};
\ee

\be \label{3.9}
\frac{\partial x^{\alpha}}{\partial \xi^{\mu}}
\frac{\partial \xi^{\mu}}{\partial x^{\beta}}
= \delta^{\alpha}_{\beta};
\ee

\be \label{3.10}
\frac{\partial ^2 t}{\partial t' \partial r'}
- \frac{\partial ^2 t}{\partial r' \partial t'}
= \frac{u u'}{1 - u^2}  ,
\ee

\be \label{3.11}
\frac{\partial ^2 t'}{\partial t \partial r}
- \frac{\partial ^2 t'}{\partial r \partial t}
= - \frac{u u'}{(1 - u^2)^{3/2}} ,
\ee
where $u' = \frac{{\rm d} u}{{\rm d} r}$.
\\

This transformation is a kind of general coordinate
transformations. Now, let's study the transformation
rules of some fundamental quantities and basic
physical equations. Under this transformation, the
metric tensors transform as:
\be \label{3.12}
\eta_{\mu\nu} \to g_{\alpha \beta}
=\frac{\partial \xi^{\mu}}{\partial x^{\alpha}}
\frac{\partial \xi^{\nu}}{\partial x^{\beta}}
\eta_{\mu\nu}
= \left (
\ba{cccc}
-1+u^2 &0&0& 0 \\
0& \frac{1}{1-u^2} &0&0  \\
0&0& r^2 &0\\
0&0&0&  r^2 {\rm sin}^2 \theta
\ea
\right ),
\ee

\be \label{3.13}
\eta^{\mu\nu} \to g^{\alpha \beta}
=\frac{\partial x^{\alpha}}{\partial \xi^{\mu}}
\frac{\partial x^{\beta}}{\partial \xi^{\nu}}
\eta^{\mu\nu}
= \left (
\ba{cccc}
- \frac{1}{1-u^2} &0&0& 0 \\
0& 1-u^2 &0&0  \\
0&0& \frac{1}{r^2} &0\\
0&0&0&  \frac{1}{r^2 {\rm sin}^2 \theta}
\ea
\right ).
\ee
The above transformation rules of metric tensors are completely
the same as those in general relativity. But the
affine connection transforms as:
\be \label{3.14}
\Gamma^{\lambda}_{\mu\nu} \to \Gamma^{\gamma}_{\alpha \beta}
= \frac{\partial \xi^{\mu}}{\partial x^{\alpha}}
\frac{\partial \xi^{\nu}}{\partial x^{\beta}}
\frac{\partial x^{\gamma}}{\partial \xi^{\lambda}}
\Gamma^{\lambda}_{\mu\nu}
+ \frac{\partial x^{\gamma}}{\partial \xi^{\mu}}
\frac{\partial ^2 \xi^{\mu}}{\partial x^{\beta}
\partial x^{\alpha}}
+ B^{\gamma}_{\alpha \beta}.
\ee
If $B^{\gamma}_{\alpha \beta}$ vanish, the transformation
rule of affine connection is completely the same as that
of general relativity. But, for the present case, it does
not vanish:
\be \label{3.15}
B^{t'}_{r' t'} =
- \frac{u u'}{(1 - u^2)^{3/2}},
\ee

\be \label{3.16}
B^{r'}_{t' t'} =
- u u' \sqrt{1 - u^2},
\ee
and other components of $B^{\gamma}_{\alpha \beta}$ vanish.
So, in the present case, the transformation rule of affine
connection is different from that of general relativity,
which will cause serious problems to the foundations
of general relativity.
Because the affine connection transforms in a different
way than that of general relativity, under this transformations,
the curvature tensor does not transforms covariantly:
\be \label{3.17}
R^{\sigma}_{\mu\nu\lambda} \to
{\mathbb R}^{\delta}_{\alpha\beta\gamma}
=\frac{\partial \xi^{\mu}}{\partial x^{\alpha}}
\frac{\partial \xi^{\nu}}{\partial x^{\beta}}
\frac{\partial \xi^{\lambda}}{\partial x^{\gamma}}
\frac{\partial x^{\delta}}{\partial \xi^{\sigma}}
R^{\sigma}_{\mu\nu\lambda}
+ G^{\delta}_{\alpha\beta\gamma},
\ee
where ${\mathbb R}^{\delta}_{\alpha\beta\gamma}$ is the curvature
tensor in coordinate system $x^{\alpha}$ and
$G^{\delta}_{\alpha\beta\gamma}$ is the non-homogeneous term which
violates covariance of curvature tensor. For the present
transformations, the non-homogeneous term
$G^{\delta}_{\alpha\beta\gamma}$ can be calculated
explicitly:
\be \label{3.18}
G^{t'}_{r' t' r'} =
- \frac{(u')^2 + u u'' - u^3 u'' }{(1-u^2)^3}
\ee

\be \label{3.19}
G^{t'}_{r' r' t'} =
 \frac{(u')^2 + u u'' - u^3 u'' }{(1-u^2)^3}
\ee

\be \label{3.20}
G^{t'}_{\theta ' t' \theta ' } =
- \frac{r u u' }{1-u^2}
\ee

\be \label{3.21}
G^{t'}_{\theta '  \theta ' t'} =
 \frac{r u u' }{1-u^2}
\ee

\be \label{3.22}
G^{t'}_{\varphi ' t' \varphi ' } =
- \frac{r u u'  {\rm sin}^2 \theta}{1-u^2}
\ee

\be \label{3.23}
G^{t'}_{\varphi '  \varphi ' t'} =
 \frac{r u u'  {\rm sin}^2 \theta}{1-u^2}
\ee

\be \label{3.24}
G^{r'}_{t' t' r'} =
- \frac{(u')^2 + u u'' - u^3 u'' }{1-u^2}
\ee

\be \label{3.25}
G^{r'}_{t' r' t' } =
 \frac{(u')^2 + u u'' - u^3 u'' }{1-u^2}
\ee

\be \label{3.26}
G^{\theta '}_{ t' t' \theta ' } =
 - \frac{u u' }{r}
\ee

\be \label{3.27}
G^{\theta '}_{ t'  \theta '  t'} =
 \frac{u u' }{r}
\ee

\be \label{3.28}
G^{\varphi '}_{ t' t' \varphi ' } =
 - \frac{u u' }{r}
\ee

\be \label{3.29}
G^{\varphi '}_{ t'  \varphi ' t' } =
 \frac{u u' }{r}
\ee
 where $u'' = \frac{{\rm d}^2 u}{{\rm d} r^2 }$.
Other component of $G^{\delta}_{\alpha \beta \gamma}$ vanish
Because the curvature tensor of flat space-time vanish
\be \label{3.30}
R^{\sigma}_{\mu\nu\lambda} = 0,
\ee
the curvature tensor ${\mathbb R}^{\delta}_{\alpha\beta\gamma}$
of the transformed space-time equals to the non-homogeneous
term $G^{\delta}_{\alpha \beta \gamma}$
\be \label{3.31}
{\mathbb R}^{\delta}_{\alpha\beta\gamma}
= G^{\delta}_{\alpha\beta\gamma}.
\ee
Eq.(\ref{3.17}) violates the transformation rules given by
general relativity. According to general relativity, because
curvature tensor is a 4th rank tensor and the curvature
tensor of flat space-time vanish, the curvature tensor
${\mathbb R}^{\delta}_{\alpha\beta\gamma}$
of the new space-time must vanish either, which
contradicts with eq.(\ref{3.31}). After index contraction,
we can obtain the transformation rule for Recci tensor
from eq.(\ref{3.17}):
\be \label{3.32}
R_{\mu\nu} \to
{\mathbb R}_{\alpha\beta}
=\frac{\partial \xi^{\mu}}{\partial x^{\alpha}}
\frac{\partial \xi^{\nu}}{\partial x^{\beta}}
R_{\mu\nu}
+ G_{\alpha\beta}
\ee
where
\be \label{3.33}
G_{\alpha \beta} = G^{\gamma}_{\alpha\gamma\beta}.
\ee
The non-vanishing component of $G_{\alpha\beta}$ are
\be \label{3.34}
G_{t' t'} =
\frac{2 u  u'}{r} +
\frac{(u')^2 + u u'' -u^3 u''}{1-u^2},
\ee

\be \label{3.35}
G_{r' r'} =
- \frac{(u')^2 + u u'' -u^3 u''}{(1-u^2)^3},
\ee

\be \label{3.36}
G_{\theta' \theta'} =
- \frac{r  u  u'}{1-u^2} ,
\ee

\be \label{3.37}
G_{\varphi' \varphi'} =
- \frac{r  u  u'  {\rm sin}^2 \theta}{1-u^2} .
\ee
Because of the existence of the non-homogeneous term
$G_{\alpha\beta}$, though the Recci tensor of flat space-time
vanishes, the Recci tensor of the new space-time
does not vanish. The corresponding transformation rule of
curvature scalar $R$ is
\be \label{3.38}
R \to {\mathbb R} = R + G,
\ee
where
\be \label{3.39}
G = g^{\alpha\beta} G_{\alpha\beta}.
\ee
In the present case, $G$ is
\be \label{3.40}
G = 2 \cdot
\frac{-2 u u' + 2 u^3 u' - r (u')^2 - r u u''
+ r u^3 u''}{r (1-u^2)^2}.
\ee
Eq.(\ref{3.38}) means that curvature scalar $R$ is no longer
a real scalar under this transformation. In other words, curvature
scalar $R$ changes its magnitude under this transformation,
or that this transformation changes the space-time
curvature; that is, it can change a flat space-time into a
curved space-time or change a curved space-time into a flat
space-time. The transformation rules of affine connection,
curvature tensor, Recci tensor and curvature scalar are
different from those in general relativity. \\

Define covariant derivatives as
\be \label{3.41}
\nabla _{\mu} A_{\nu} =
\frac{\partial A_{\nu}}{\partial \xi^{\mu}}
- \Gamma^{\lambda}_{\nu \mu} A_{\lambda},
\ee

\be \label{3.42}
\nabla _{\mu} A^{\nu} =
\frac{\partial A^{\nu}}{\partial \xi^{\mu}}
+ \Gamma^{\nu}_{\mu \lambda} A^{\lambda}.
\ee
Under this transformation, the transformation rules of
covariant derivatives are
\be \label{3.43}
\nabla_{\mu} A_{\nu} \to \nabla_{\alpha} A_{\beta}
= \frac{\partial \xi^{\mu}}{\partial x^{\alpha}}
\frac{\partial \xi^{\nu}}{\partial x^{\beta}}
\nabla_{\mu} A_{\nu}
- B^{\gamma}_{\beta\alpha}
\frac{\partial \xi^{\nu}}{\partial x^{\gamma}} A_{\nu},
\ee

\be \label{3.44}
\nabla_{\mu} A^{\nu} \to \nabla_{\alpha} A^{\beta}
= \frac{\partial \xi^{\mu}}{\partial x^{\alpha}}
\frac{\partial x^{\beta}}{\partial \xi^{\nu}}
\nabla_{\mu} A^{\nu}
+ B^{\beta}_{\alpha\gamma}
\frac{\partial x^{\gamma}}{\partial \xi^{\nu}} A^{\nu}
+ \frac{\partial \xi^{\mu}}{\partial x^{\alpha}}
\left (\frac{\partial^2 x^{\beta}}{\partial \xi^{\mu}
\partial \xi^{\nu}}
- \frac{\partial^2 x^{\beta}}{\partial \xi^{\nu}
\partial \xi^{\mu}} \right ) A^{\nu}.
\ee
Therefore, the covariant derivatives are not a real covariant
under this transformation. It can be proved that
\be \label{3.45}
\nabla_{t'} A_{r'} =
\frac{\partial \xi^{\mu}}{\partial t'}
\frac{\partial \xi^{\nu}}{\partial r'}
\nabla_{\mu} A_{\nu}
+ \frac{u u'}{1-u^2} A_t,
\ee

\be \label{3.46}
\nabla_{t'} A_{t'} =
\frac{\partial \xi^{\mu}}{\partial t'}
\frac{\partial \xi^{\nu}}{\partial t'}
\nabla_{\mu} A_{\nu}
+ u u' A_r,
\ee

\be \label{3.47}
\nabla_{t'} A^{r'} =
\frac{\partial \xi^{\mu}}{\partial t'}
\frac{\partial r'}{\partial \xi^{\nu}}
\nabla_{\mu} A^{\nu}
- u u'  A^t
\ee

\be \label{3.48}
\nabla_{t'} A^{t'} =
\frac{\partial \xi^{\mu}}{\partial t'}
\frac{\partial t'}{\partial \xi^{\nu}}
\nabla_{\mu} A^{\nu}
- \frac{u u'}{1-u^2}  A^r,
\ee
and other components of covariant derivative transform
covariantly. Therefore, under the transformation eq.(\ref{3.2}),
the conventional covariant derivatives defined in general
relativity do not transform covariantly, which will cause
serious problems to the foundations of general
relativity. \\

\section{Transformation Rules of Basic Physical Equations}
\setcounter{equation}{0}

Now, let's discuss the transformation rules of basic physical
equations. It is generally believed that, in general relativity,
all basic physical equations should transform covariantly
under a general coordinate transformation. Here, we
directly calculate the transformation rule of these basic physical
equations eqs.(\ref{2.5} - \ref{2.8}) and eq.(\ref{2.10})  to
see whether they transform covariantly. Without losing generality,
suppose that the basic physical equations in reference system
$\xi^{\mu}$ are given by  eqs.(\ref{2.5} - \ref{2.8}) and eq.(\ref{2.10}).
We will directly calculate the corresponding physical equations
in the new coordinate system $x^{\alpha}$. \\

Suppose that the conserved current $J^{\mu}$ transforms covariantly
under the transformation eq.(\ref{3.2}), then eq.(\ref{2.5}) will
be transformed into the following form
\be \label{4.1}
\nabla_{\alpha} J^{\alpha}
=  B^{\alpha}_{\alpha\gamma}
\frac{\partial x^{\gamma}}{\partial \xi^{\nu}} J^{\nu}
+ \frac{\partial \xi^{\mu}}{\partial x^{\alpha}}
\left (\frac{\partial^2 x^{\alpha}}{\partial \xi^{\mu}
\partial \xi^{\nu}}
- \frac{\partial^2 x^{\alpha}}{\partial \xi^{\nu}
\partial \xi^{\mu}} \right ) J^{\nu}.
\ee
It can be written out explicitly
\be \label{4.2}
\nabla_{\alpha} J^{\alpha}
=  - \frac{u u'}{1-u^2} J^r.
\ee
Because the right hand side of eq.(\ref{4.2}) does not vanish,
current $J^{\alpha}$ is not a conserved current in the new
coordinate system $x^{\alpha}$, or in the new coordinate
system, the continuity equation does not hold. The reason
is simple: the transformation matrices eqs.(\ref{3.4} - \ref{3.5})
or new coordiantes $x^{\alpha}$ carry some dynamics. The space-time
itself carries some dynamics, the physical vacuum of a physical
system is not a conserved system, so the physical system alone
is not a conserved system. The unite system of physical vacuum
and physical system is a conserved system. \\

Under the transformation eq.(\ref{3.2}),
the geodesic equation eq.(\ref{2.6}) is changed into
\be \label{4.3}
\frac{{\rm d}^2 x^{\alpha}}{{\rm d} \tau^2}
+ \Gamma^{\alpha}_{\beta \gamma}
\frac{{\rm d} x^{\beta}}{{\rm d} \tau}
\frac{{\rm d} x^{\gamma}}{{\rm d} \tau}
-B^{\alpha}_{\beta \gamma}
\frac{\partial x^{\beta}}{\partial \xi^{\nu}}
\frac{\partial x^{\gamma}}{\partial \xi^{\lambda}}
\frac{{\rm d}\xi^{\nu}}{{\rm d} \tau}
\frac{{\rm d}\xi^{\lambda}}{{\rm d} \tau} =0.
\ee
This equation can be written out explicitly
\be \label{4.4}
\frac{{\rm d}^2 t'}{{\rm d} \tau^2}
+ \Gamma^{t'}_{\beta \gamma}
\frac{{\rm d} x^{\beta}}{{\rm d} \tau}
\frac{{\rm d} x^{\gamma}}{{\rm d} \tau}
= - \frac{u u'}{(1-u^2)^{3/2}}
\frac{{\rm d}t}{{\rm d} \tau}
\frac{{\rm d}r}{{\rm d} \tau},
\ee

\be \label{4.5}
\frac{{\rm d}^2 r'}{{\rm d} \tau^2}
+ \Gamma^{r'}_{\beta \gamma}
\frac{{\rm d} x^{\beta}}{{\rm d} \tau}
\frac{{\rm d} x^{\gamma}}{{\rm d} \tau}
= - \frac{u u'}{\sqrt{1-u^2}}
\left (\frac{{\rm d}t}{{\rm d} \tau}\right )^2,
\ee

\be \label{4.6}
\frac{{\rm d}^2 \theta'}{{\rm d} \tau^2}
+ \Gamma^{ \theta'}_{\beta \gamma}
\frac{{\rm d} x^{\beta}}{{\rm d} \tau}
\frac{{\rm d} x^{\gamma}}{{\rm d} \tau}
= 0,
\ee

\be \label{4.7}
\frac{{\rm d}^2 \varphi'}{{\rm d} \tau^2}
+ \Gamma^{ \varphi'}_{\beta \gamma}
\frac{{\rm d} x^{\beta}}{{\rm d} \tau}
\frac{{\rm d} x^{\gamma}}{{\rm d} \tau}
= 0.
\ee
Because the right hand sides of eqs.(\ref{4.4} -- \ref{4.5})
do not vanish, the geodesic equation is not satisfied in
the new coordinate system $x^{\alpha}$. Or in other words,
if the mass point moves along the geodesic line in the original
reference system $\xi^{\mu}$, it will not move along the geodesic
line in the new coordinate system $x^{\alpha}$.
It means that, in an arbitrary curved space-time and in an arbitrary
coordinate system, a mass point will not move along the geodesic line,
or in most general case, the geodesic line is not the trajectory
of a free mass point.  \\

According to general relativity, the parallel transport equation
in an arbitrary curved space-time is given by eq.(\ref{2.7}). Supposed
that in the reference system $\xi^{\mu}$, the parallel transport
equation is satisfied. After the transformation eq.(\ref{3.2}),
it will be changed into
\be \label{4.8}
\frac{{\rm D} A^{\alpha}}{{\rm D} \tau} =
 \frac{\partial x^{\beta}}{\partial \xi^{\nu}}
\frac{\partial x^{\gamma}}{\partial \xi^{\lambda}}
B^{\alpha}_{\beta \gamma}
\frac{{\rm d} \xi^{\lambda}}{{\rm d} \tau} A^{\nu}.
\ee
Its explicit form is
\be \label{4.9}
\frac{{\rm D} A^{t'}}{{\rm D} \tau} =
- \frac{u u'}{(1-u^2)^{3/2}}
\frac{{\rm d}t}{{\rm d} \tau} A^r,
\ee

\be \label{4.10}
\frac{{\rm D} A^{r'}}{{\rm D} \tau} =
- \frac{u u'}{\sqrt{1-u^2}}
\frac{{\rm d}t}{{\rm d} \tau} A^t,
\ee

\be \label{4.11}
\frac{{\rm D} A^{ \theta'}}{{\rm D} \tau} = 0,
\ee

\be \label{4.12}
\frac{{\rm D} A^{ \varphi'}}{{\rm D} \tau} = 0.
\ee
For a covariant vector $B_{\mu}$, according to general
relativity, its parallel transport equation is
\be \label{4.13}
\frac{{\rm D} B_{\mu}}{{\rm D} \tau} = 0,
\ee
where
\be \label{4.14}
\frac{{\rm D} B_{\mu}}{{\rm D} \tau} =
\frac{{\rm d} B_{\mu}}{{\rm d} \tau}
- \Gamma^{\lambda}_{\mu \nu}
\frac{{\rm d} \xi^{\nu}}{{\rm d} \tau}
B_{\lambda}.
\ee
Suppose that in the original reference system, covariant vector
$B_{\mu}$ satisfies parallel transport equation eq.(\ref{4.13}).
Then after the transformation eq.(\ref{3.2}), it will become
\be \label{4.15}
\frac{{\rm D} B_{\alpha}}{{\rm D} \tau} =
- \frac{\partial x^{\beta}}{\partial \xi^{\nu}}
\frac{\partial \xi^{\lambda}}{\partial x^{\gamma}}
B^{\gamma}_{\alpha \beta }
\frac{{\rm d} \xi^{\nu}}{{\rm d} \tau} B_{\lambda}.
\ee
It can be written out explicitly
\be \label{4.16}
\frac{{\rm D} B_{t'}}{{\rm D} \tau} =
\frac{u u'}{\sqrt{1-u^2}}
\frac{{\rm d}t}{{\rm d} \tau} B_r,
\ee

\be \label{4.17}
\frac{{\rm D} B_{r'}}{{\rm D} \tau} =
\frac{u u'}{(1-u^2)^{3/2}}
\frac{{\rm d}t}{{\rm d} \tau} B_t,
\ee

\be \label{4.18}
\frac{{\rm D} B_{\theta'}}{{\rm D} \tau} = 0,
\ee

\be \label{4.19}
\frac{{\rm D} B_{\varphi'}}{{\rm D} \tau} = 0.
\ee
Because the right hand side of eqs.(\ref{4.9} -- \ref{4.10})
and eqs.(\ref{4.16} -- \ref{4.17}) do not vanish, the parallel
transport equations do not hold in the new coordinate system.\\

In general relativity, the energy-momentum conservation equation
is given by eq.(\ref{2.8}). Suppose that it holds in the reference
system $\xi^{\mu}$. Then the transformation eq.(\ref{3.2}) will
make it change into
\be \label{4.20}
\nabla_{\alpha} T^{\beta \alpha}  =
\frac{\partial \xi^{\mu}}{\partial x^{\alpha}}
\frac{\partial x^{\beta}}{\partial \xi^{\nu}}
\left (
\frac{\partial^2 x^{\alpha}}{\partial \xi^{\mu} \partial \xi^{\lambda}}
-\frac{\partial^2 x^{\alpha}}{\partial \xi^{\lambda} \partial \xi^{\mu}}
\right ) T^{\nu \lambda}
 + \frac{\partial x^{\alpha}}{\partial \xi^{\lambda}}
\frac{\partial x^{\delta}}{\partial \xi^{\sigma}}
B^{\beta}_{\alpha \delta} T^{\sigma \lambda}
+\frac{\partial x^{\beta}}{\partial \xi^{\nu}}
\frac{\partial x^{\delta}}{\partial \xi^{\sigma}}
B^{\alpha}_{\alpha \delta} T^{\nu \sigma}.
\ee
Its explicit form is
\be \label{4.21}
\nabla_{\alpha} T^{t' \alpha}  =
- \frac{2 u u'}{(1-u^2)^{3/2}} T^{tr},
\ee

\be \label{4.22}
\nabla_{\alpha} T^{r' \alpha}  =
- \frac{ u u'}{\sqrt{1-u^2}} (T^{tt}+T^{rr}) ,
\ee

\be \label{4.23}
\nabla_{\alpha} T^{\theta' \alpha}  =
- \frac{ u u'}{1-u^2} T^{\theta r},
\ee

\be \label{4.24}
\nabla_{\alpha} T^{\varphi' \alpha}  =
- \frac{ u u'}{1-u^2} T^{\varphi r}.
\ee
These equations mean that, in the new coordinate system
$x^{\alpha}$, energy-momentum is no longer conserved.
This result is a little surprising, but is easily to understand.
Suppose that the reference system $\xi^{\mu}$ is a local inertial
reference system and a mass point moves freely in it. Because of
energy-momentum conservation, the speed of the mass point will not
change with time. But in the new coordinate system $x^{\alpha}$,
because of its variable motion, the speed of the mass point will
change with time. So, the observer in the new coordinate system
$x^{\alpha}$ will find that the energy-momentum of the mass point
changes with time, or in other words, its energy-momentum is
not conserved. \\

Finally, let's discuss Einstein field equation eq.(\ref{2.10}).
We also suppose that it holds in the original reference ssytem
$\xi^{\mu}$. Transformation eq.(\ref{3.2}) changes it into
\be  \label{4.25}
{\mathbb R}_{\alpha \beta}
-  \frac{1}{2} g_{\alpha \beta} {\mathbb R}
+ 8 \pi G_N T_{\alpha \beta}
= G_{\alpha \beta} -  \frac{1}{2} g_{\alpha \beta} G.
\ee
If the right hand side of the above equation vanish, the Einstein
field equation still holds in the new coordinate system $x^{\alpha}$.
However, it does not vanish. In the new coordinate system
$x^{\alpha}$, the diagonal component equations are
\be  \label{4.26}
{\mathbb R}_{t' t'}
-  \frac{1}{2} g_{t' t'} {\mathbb R}
+ 8 \pi G_N T_{t' t'}
= 0,
\ee

\be  \label{4.27}
{\mathbb R}_{r' r'}
-  \frac{1}{2} g_{r' r'} {\mathbb R}
+ 8 \pi G_N T_{r' r'}
= \frac{2 u u'}{r (1 - u^2)^2},
\ee

\be  \label{4.28}
{\mathbb R}_{\theta' \theta'}
-  \frac{1}{2} g_{\theta' \theta'} {\mathbb R}
+ 8 \pi G_N T_{\theta' \theta'}
= \frac{r u u' - r u^3 u' + r^2 (u')^2
+ r^2 u u'' - r^2 u^3 u''}
{ (1 - u^2)^2},
\ee

\be  \label{4.29}
{\mathbb R}_{\varphi' \varphi'}
-  \frac{1}{2} g_{\varphi' \varphi'} {\mathbb R}
+ 8 \pi G_N T_{\varphi' \varphi'}
= \frac{r u u' - r u^3 u' + r^2 (u')^2
+ r^2 u u'' - r^2 u^3 u''}
{ (1 - u^2)^2} {\rm sin}^2 \theta.
\ee
Therefore, it is clear that Einstein field equation does
not transform covriantly under a very simple coordinate
transformation eq.(\ref{3.2}). This result is obtained by
direct calculation. Einstein field equation is not
general covariant. It means that, if in the original reference
system $\xi^{\mu}$, the Einstein field equation is
satisfied, it will not be satisfied in the new coordinate
system $x^{\alpha}$. So, if an observer in an arbitrary
curved space-time, he will not know that that whether
the Einstein  field equation is satisfied or not in his
coordinate system, or in other words, Einstein field equation
does not hold in an arbitrary coordinate system of an arbitrary
curved space-time.

\section{A General theory on General Coordinate Transformation}
\setcounter{equation}{0}

In the chapter 4, through a simple example, we know that basic
physical equations are not covariant under a general coordinate
transformation, because almost all physical equations change their
forms under the general coordinate
transformation eq.(\ref{3.2}). But in general
relativity, it is also proved that all physical equations preserve
their forms under a general coordinate transformation. These two results
are contradicted each other. In this chapter, we will develop a general
theory on general coordinate transformations, and study what cause
the violation of  general covariance in general relativity.  \\

First, we need to point out that transformation eq.(\ref{3.2}) is a kind
of general coordinate transformations. It is a simple example of
general coordinate transformations. Essentially speaking, local Lorentz
transformation belongs to the same kind of general coordinate
transformations\cite{wu8}. For a general coordinate transformation,
\be \label{5.2}
{\rm d} \xi^{\mu} \to {\rm d} x^{\alpha}
= \frac{\partial x^{\alpha}}{\partial \xi^{\mu}}
{\rm d} \xi^{\mu},
\ee

\be \label{5.3}
\frac{\partial}{ \partial \xi^{\mu}}  \to
\frac{\partial}{\partial x^{\alpha}} =
\frac{\partial \xi^{\mu}}{\partial x^{\alpha}}
\frac{\partial}{\partial \xi^{\mu}},
\ee
the transformation matrix must satisfy the following restriction:
\be \label{5.4}
\frac{\partial \xi^{\mu}}{\partial x^{\alpha}}
\frac{\partial x^{\alpha}}{\partial \xi^{\nu}}
= \delta^{\mu}_{\nu};
\ee

\be \label{5.5}
\frac{\partial x^{\alpha}}{\partial \xi^{\mu}}
\frac{\partial \xi^{\mu}}{\partial x^{\beta}}
= \delta^{\alpha}_{\beta}.
\ee
There are two kinds of general coordinate transformations. The first
kind is a trivial general coordinate transformation, or call it
curvilinear coordinate transformation. Under this transformation,
the space-time itself undergoes no changes. The only change is that we
transform one curvilinear coordinate system into another curvilinear
coordinate system. For curvilinear coordinate transformation, the
transformation matrices satisfy:
\be \label{5.6}
\frac{\partial ^2 \xi^{\mu}}{\partial x^{\alpha} \partial x^{\beta}}
- \frac{\partial ^2 \xi^{\mu}}{\partial x^{\beta} \partial x^{\alpha}}
= 0 ,
\ee

\be \label{5.7}
\frac{\partial ^2 x^{\alpha}}{\partial \xi^{\mu} \partial \xi^{\nu}}
- \frac{\partial ^2 x^{\alpha}}{\partial \xi^{\nu} \partial \xi^{\mu}}
= 0.
\ee
The transformation between Cartesian coordinate system and
spherical coordinate system belongs to this kind.
Both transformation eq.(\ref{3.2}) and local Lorentz transformation
do not satisfy the above  two restrictions eqs.(\ref{5.6} -- \ref{5.7}),
so they do not belong to this kind.
\\

For the second kind of general coordinate transformations, the transformation
matrix do not satisfy the above two restrictions
eqs.(\ref{5.6} -- \ref{5.7}), i.e.
\be \label{5.8}
\frac{\partial ^2 \xi^{\mu}}{\partial x^{\alpha} \partial x^{\beta}}
- \frac{\partial ^2 \xi^{\mu}}{\partial x^{\beta} \partial x^{\alpha}}
\not= 0 ,
\ee

\be \label{5.9}
\frac{\partial ^2 x^{\alpha}}{\partial \xi^{\mu} \partial \xi^{\nu}}
- \frac{\partial ^2 x^{\alpha}}{\partial \xi^{\nu} \partial \xi^{\mu}}
\not= 0.
\ee
A lot of important general coordinate transformations, such as
transformation eq.(\ref{3.2}) and local Lorentz transformation,
belong to this kind of general coordinate transformations. On of
the most important transformation in general relativity, the
transformation from local inertial reference system to
an arbitrary curved space-time, belongs to this kind. In this
chapter, we mainly discuss this kind of general coordinate
transformations. Later we will know that the violation of
eqs(\ref{5.6} -- \ref{5.7}) causes the violation of  general
covariance in general relativity.  \\

The general coordinate system before the transformation is
denoted as $\xi^{\mu}$ (please note that in this chapter,
$\xi^{\mu}$ is no longer coordinate system of flat space-time.)
and that after the transformation is denoted as $x^{\alpha}$.
The transformation from one general coordinate system
$\xi^{\mu}$ to another general coordinate system $x^{\alpha}$
is
\be \label{5.11}
{\rm d} \xi^{\mu} \to {\rm d} x^{\alpha}
= \frac{\partial x^{\alpha}}{\partial \xi^{\mu}}
{\rm d} \xi^{\mu},
\ee

\be \label{5.12}
\frac{\partial}{ \partial \xi^{\mu}}  \to
\frac{\partial}{\partial x^{\alpha}} =
\frac{\partial \xi^{\mu}}{\partial x^{\alpha}}
\frac{\partial}{\partial \xi^{\mu}},
\ee
where transformation matrices
$\frac{\partial x^{\alpha}}{\partial \xi^{\mu}}$
and
$\frac{\partial \xi^{\mu}}{\partial x^{\alpha}}$
satisfy eqs.(\ref{3.8} -- \ref{3.9}) and
eqs(\ref{5.8} -- \ref{5.9}).
Under this transformation, the
metric tensors transform as:
\be \label{5.13}
\eta_{\mu\nu} \to g_{\alpha \beta}
=\frac{\partial \xi^{\mu}}{\partial x^{\alpha}}
\frac{\partial \xi^{\nu}}{\partial x^{\beta}}
\eta_{\mu\nu},
\ee

\be \label{5.14}
\eta^{\mu\nu} \to g^{\alpha \beta}
=\frac{\partial x^{\alpha}}{\partial \xi^{\mu}}
\frac{\partial x^{\beta}}{\partial \xi^{\nu}}
\eta^{\mu\nu},
\ee
where $\eta^{\mu \nu}$ is the metric tensor of the coordinate
system $\xi^{\mu}$( In this chapter, $\eta^{\mu\nu}$ is not
Minkowski metric.).
The affine connection transforms as:
\be \label{5.15}
\Gamma^{\lambda}_{\mu\nu} \to \Gamma^{\gamma}_{\alpha \beta}
= \frac{\partial \xi^{\mu}}{\partial x^{\alpha}}
\frac{\partial \xi^{\nu}}{\partial x^{\beta}}
\frac{\partial x^{\gamma}}{\partial \xi^{\lambda}}
\Gamma^{\lambda}_{\mu\nu}
+ \frac{\partial x^{\gamma}}{\partial \xi^{\mu}}
\frac{\partial ^2 \xi^{\mu}}{\partial x^{\beta}
\partial x^{\alpha}}
+ B^{\gamma}_{\alpha \beta},
\ee
where
\be \label{5.16}
\ba{rcl}
B^{\gamma}_{\alpha \beta}
& = & \frac{1}{2}
\frac{\partial x^{\gamma}}{\partial \xi^{\mu}}
\left (
\frac{\partial ^2 \xi^{\mu}}{\partial x^{\alpha}
\partial x^{\beta}}
-\frac{\partial ^2 \xi^{\mu}}{\partial x^{\beta}
\partial x^{\alpha}}
\right)  \\
&&\\
&& + \frac{1}{2}
\frac{\partial \xi^{\mu}}{\partial x^{\alpha}}
\frac{\partial x^{\gamma}}{\partial \xi^{\lambda}}
\frac{\partial x^{\delta}}{\partial \xi^{\sigma_1}}
\eta_{\mu \nu} \eta^{\lambda \sigma_1}
\left (
\frac{\partial ^2 \xi^{\nu}}{\partial x^{\beta}
\partial x^{\delta}}
-\frac{\partial ^2 \xi^{\nu}}{\partial x^{\delta}
\partial x^{\beta}}
\right) \\
&&\\
&& + \frac{1}{2}
\frac{\partial \xi^{\nu}}{\partial x^{\beta}}
\frac{\partial x^{\gamma}}{\partial \xi^{\lambda}}
\frac{\partial x^{\delta}}{\partial \xi^{\sigma_1}}
\eta_{\mu \nu} \eta^{\lambda \sigma_1}
\left (
\frac{\partial ^2 \xi^{\mu}}{\partial x^{\alpha}
\partial x^{\delta}}
-\frac{\partial ^2 \xi^{\mu}}{\partial x^{\delta}
\partial x^{\alpha}}
\right).
\ea
\ee
For the first kind of general coordinate transformations,
eqs.(\ref{5.6} -- \ref{5.7}) are satisfied, so
$B^{\gamma}_{\alpha \beta}$ vanishes and the transformation
rule of affine connection becomes the same as that of
general relativity. For the second kind of general coordinate
transformations, $B^{\gamma}_{\alpha \beta}$ does not vanish and
the transformation rule of affine connection is different that of
general relativity. The appearance of the non-homogeneous term
$B^{\gamma}_{\alpha \beta}$  is the origin of the violation of
general covariance of general relativity, and the violation of
eqs.(\ref{5.6} -- \ref{5.7}) is the origin of the appearance
of $B^{\gamma}_{\alpha \beta}$, so the violation of
eqs.(\ref{5.6} -- \ref{5.7}) is the essential  origin
of the violation of general covariance of general relativity. \\

Under the second kind of general coordinate transformations, curvature
tensor transforms as
\be \label{5.17}
R^{\sigma}_{\mu\nu\lambda} \to
{\mathbb R}^{\delta}_{\alpha\beta\gamma}
=\frac{\partial \xi^{\mu}}{\partial x^{\alpha}}
\frac{\partial \xi^{\nu}}{\partial x^{\beta}}
\frac{\partial \xi^{\lambda}}{\partial x^{\gamma}}
\frac{\partial x^{\delta}}{\partial \xi^{\sigma}}
R^{\sigma}_{\mu\nu\lambda}
+ G^{\delta}_{\alpha\beta\gamma},
\ee
where ${\mathbb R}^{\delta}_{\alpha\beta\gamma}$ is the curvature
tensor in coordinate system $x^{\alpha}$ and
$G^{\delta}_{\alpha\beta\gamma}$ is the non-homogeneous term which
violates general covariance of curvature tensor. Explicit form of
$G^{\delta}_{\alpha\beta\gamma}$ is quite complicated
\be \label{5.18}
G^{\delta}_{\alpha\beta\gamma}
= \sum_{i=1}^{8} G^{\delta}_{(i) \alpha\beta\gamma},
\ee
where
\be \label{5.19}
\ba{rcl}
G^{\delta}_{(1)\alpha\beta\gamma}
& = & \frac{1}{2}
\frac{\partial \xi^{\nu}}{\partial x^{\beta}}
\frac{\partial x^{\delta}}{\partial \xi^{\kappa}}
\left (
\frac{\partial ^2 \xi^{\mu}}{\partial x^{\gamma}
\partial x^{\alpha}}
-\frac{\partial ^2 \xi^{\mu}}{\partial x^{\alpha}
\partial x^{\gamma}}
\right)  \Gamma^{\kappa}_{\mu \nu}
- ( \beta \leftrightarrow \gamma)\\
&&\\
&& + \frac{\partial \xi^{\mu}}{\partial x^{\alpha}}
\frac{\partial x^{\delta}}{\partial \xi^{\kappa}}
\left (
\frac{\partial ^2 \xi^{\nu}}{\partial x^{\gamma}
\partial x^{\beta}}
-\frac{\partial ^2 \xi^{\nu}}{\partial x^{\beta}
\partial x^{\gamma}}
\right)  \Gamma^{\kappa}_{\mu \nu},
\ea
\ee

\be \label{5.20}
\ba{rcl}
G^{\delta}_{(2)\alpha\beta\gamma}
& = & \frac{\partial \xi^{\mu}}{\partial x^{\alpha}}
\frac{\partial \xi^{\nu}}{\partial x^{\beta}}
\frac{\partial \xi^{\lambda}}{\partial x^{\gamma}}
\left (
\frac{\partial ^2 x^{\delta}}{\partial \xi^{\lambda}
\partial \xi^{\kappa}}
-\frac{\partial ^2 x^{\delta}}{\partial \xi^{\kappa}
\partial \xi^{\lambda}}
\right)  \Gamma^{\kappa}_{\mu \nu}
- ( \beta \leftrightarrow \gamma)\\
&&\\
&&
+ \frac{1}{2}
\frac{\partial \xi^{\mu}}{\partial x^{\alpha}}
\frac{\partial \xi^{\nu}}{\partial x^{\beta}}
\frac{\partial x^{\alpha_1}}{\partial \xi^{\mu_1}}
\frac{\partial x^{\delta}}{\partial \xi^{\kappa}}
\left (
\frac{\partial ^2 \xi^{\kappa}}{\partial x^{\gamma}
\partial x^{\alpha_1}}
-\frac{\partial ^2 \xi^{\kappa}}{\partial x^{\alpha_1}
\partial x^{\gamma}}
\right)  \Gamma^{\mu_1}_{\mu \nu}
- ( \beta \leftrightarrow \gamma),
\ea
\ee

\be \label{5.21}
\ba{rcl}
G^{\delta}_{(3)\alpha\beta\gamma}
& = & \frac{1}{2}
\frac{\partial \xi^{\mu}}{\partial x^{\alpha}}
\frac{\partial \xi^{\nu}}{\partial x^{\beta}}
\frac{\partial \xi^{\lambda}}{\partial x^{\gamma}}
\frac{\partial x^{\alpha_1}}{\partial \xi^{\mu_1}}
g^{\delta \varepsilon} \eta_{\lambda \mu_2}
\left (
\frac{\partial ^2 \xi^{\mu_2}}{\partial x^{\alpha_1}
\partial x^{\varepsilon}}
-\frac{\partial ^2 \xi^{\mu_2}}{\partial x^{\varepsilon}
\partial x^{\alpha_1}}
\right)  \Gamma^{\mu_1}_{\mu \nu}
- ( \beta \leftrightarrow \gamma)\\
&&\\
&&
+\frac{1}{2}
\frac{\partial \xi^{\mu}}{\partial x^{\alpha}}
\frac{\partial \xi^{\nu}}{\partial x^{\beta}}
g^{\delta \varepsilon} \eta_{\lambda \mu_1}
\left (
\frac{\partial ^2 \xi^{\lambda}}{\partial x^{\gamma}
\partial x^{\varepsilon}}
-\frac{\partial ^2 \xi^{\lambda}}{\partial x^{\varepsilon}
\partial x^{\gamma}}
\right)  \Gamma^{\mu_1}_{\mu \nu}
- ( \beta \leftrightarrow \gamma)\\
&&\\
&& +  \frac{1}{2}
\frac{\partial \xi^{\mu}}{\partial x^{\alpha}}
\frac{\partial \xi^{\mu_1}}{\partial x^{\alpha_1}}
\frac{\partial \xi^{\lambda}}{\partial x^{\gamma}}
\frac{\partial x^{\delta}}{\partial \xi^{\kappa}}
g^{\alpha_1 \varepsilon} \eta_{\mu \nu}
\left (
\frac{\partial ^2 \xi^{\nu}}{\partial x^{\beta}
\partial x^{\varepsilon}}
-\frac{\partial ^2 \xi^{\nu}}{\partial x^{\varepsilon}
\partial x^{\beta}}
\right)  \Gamma^{\kappa}_{\lambda \mu_1 }
- ( \beta \leftrightarrow \gamma)\\
&&\\
&& + \frac{1}{2}
\frac{\partial \xi^{\nu}}{\partial x^{\beta}}
\frac{\partial \xi^{\mu_1}}{\partial x^{\alpha_1}}
\frac{\partial \xi^{\lambda}}{\partial x^{\gamma}}
\frac{\partial x^{\delta}}{\partial \xi^{\kappa}}
g^{\alpha_1 \varepsilon} \eta_{\mu \nu}
\left (
\frac{\partial ^2 \xi^{\mu}}{\partial x^{\alpha}
\partial x^{\varepsilon}}
-\frac{\partial ^2 \xi^{\mu}}{\partial x^{\varepsilon}
\partial x^{\alpha}}
\right)  \Gamma^{\kappa}_{\lambda \mu_1 }
- ( \beta \leftrightarrow \gamma),
\ea
\ee

\be \label{5.22}
\ba{rcl}
G^{\delta}_{(4)\alpha\beta\gamma}
& = & \frac{1}{2}
\frac{\partial \xi^{\nu}}{\partial x^{\beta}}
g^{\delta \varepsilon} \eta_{\mu \nu}
\frac{\partial }{\partial x^{\gamma}}
\left (
\frac{\partial ^2 \xi^{\mu}}{\partial x^{\alpha}
\partial x^{\varepsilon}}
-\frac{\partial ^2 \xi^{\mu}}{\partial x^{\varepsilon}
\partial x^{\alpha}}
\right)  - ( \beta \leftrightarrow \gamma)\\
&&\\
&&
+ \frac{1}{2}
\frac{\partial \xi^{\mu}}{\partial x^{\alpha}}
g^{\delta \varepsilon} \eta_{\mu \nu}
\frac{\partial }{\partial x^{\gamma}}
\left (
\frac{\partial ^2 \xi^{\nu}}{\partial x^{\beta}
\partial x^{\varepsilon}}
-\frac{\partial ^2 \xi^{\nu}}{\partial x^{\varepsilon}
\partial x^{\beta}}
\right)  - ( \beta \leftrightarrow \gamma)\\
&&\\
&& + \frac{1}{2}
\frac{\partial x^{\delta}}{\partial \xi^{\kappa}}
\left (
\frac{\partial ^3 \xi^{\kappa}}{\partial x^{\gamma}
\partial x^{\alpha} \partial x^{\beta}}
+ \frac{\partial ^3 \xi^{\kappa}}{\partial x^{\gamma}
\partial x^{\beta} \partial x^{\alpha}}
- \frac{\partial ^3 \xi^{\kappa}}{\partial x^{\beta}
\partial x^{\alpha} \partial x^{\gamma}}
- \frac{\partial ^3 \xi^{\kappa}}{\partial x^{\beta}
\partial x^{\gamma} \partial x^{\alpha}}
\right) ,
\ea
\ee

\be \label{5.23}
\ba{rcl}
G^{\delta}_{(5)\alpha\beta\gamma}
& = & \frac{1}{2}
\frac{\partial \xi^{\lambda}}{\partial x^{\gamma}}
\left (
\frac{\partial ^2 x^{\delta}}{\partial \xi^{\lambda}
\partial \xi^{\kappa}}
-\frac{\partial ^2 x^{\delta}}{\partial \xi^{\kappa}
\partial \xi^{\lambda}}
\right)  \left (
\frac{\partial ^2 \xi^{\kappa}}{\partial x^{\alpha}
\partial x^{\beta}}
+ \frac{\partial ^2 \xi^{\kappa}}{\partial x^{\beta}
\partial x^{\alpha}}
\right)  - ( \beta \leftrightarrow \gamma)\\
&&\\
&&
+ \frac{1}{4}
\frac{\partial x^{\alpha_1}}{\partial \xi^{\mu_1}}
\frac{\partial x^{\delta}}{\partial \xi^{\kappa}}
\left (
\frac{\partial ^2 \xi^{\kappa}}{\partial x^{\gamma}
\partial x^{\alpha_1}}
-\frac{\partial ^2 \xi^{\kappa}}{\partial x^{\alpha_1}
\partial x^{\gamma}}
\right)  \left (
\frac{\partial ^2 \xi^{\mu_1}}{\partial x^{\alpha}
\partial x^{\beta}}
+ \frac{\partial ^2 \xi^{\mu_1}}{\partial x^{\beta}
\partial x^{\alpha}}
\right) - ( \beta \leftrightarrow \gamma),
\ea
\ee

\be \label{5.24}
\ba{rcl}
G^{\delta}_{(6)\alpha\beta\gamma}
& = & \frac{1}{2} \left \lbrack
\frac{\partial }{\partial x^{\gamma}}
\left (
\frac{\partial \xi^{\mu}}{\partial x^{\alpha}}
g^{\delta \varepsilon} \eta_{\mu \nu}
\right ) \right \rbrack
\left (
\frac{\partial ^2 \xi^{\nu}}{\partial x^{\beta}
\partial x^{\varepsilon}}
-\frac{\partial ^2 \xi^{\nu}}{\partial x^{\varepsilon}
\partial x^{\beta}}
\right)  - ( \beta \leftrightarrow \gamma)\\
&&\\
&&
+ \frac{1}{2} \left \lbrack
\frac{\partial }{\partial x^{\gamma}}
\left (
\frac{\partial \xi^{\nu}}{\partial x^{\beta}}
g^{\delta \varepsilon} \eta_{\mu \nu}
\right ) \right \rbrack
\left (
\frac{\partial ^2 \xi^{\mu}}{\partial x^{\alpha}
\partial x^{\varepsilon}}
-\frac{\partial ^2 \xi^{\mu}}{\partial x^{\varepsilon}
\partial x^{\alpha}}
\right)  - ( \beta \leftrightarrow \gamma),
\ea
\ee

\be \label{5.25}
\ba{rcl}
G^{\delta}_{(7)\alpha\beta\gamma}
& = & \frac{1}{4}
\frac{\partial x^{\alpha_1}}{\partial \xi^{\mu_1}}
\frac{\partial \xi^{\lambda}}{\partial x^{\gamma}}
g^{\delta \varepsilon} \eta_{\lambda \mu_2}
\left (
\frac{\partial ^2 \xi^{\mu_2}}{\partial x^{\alpha_1}
\partial x^{\varepsilon}}
-\frac{\partial ^2 \xi^{\mu_2}}{\partial x^{\varepsilon}
\partial x^{\alpha_1}}
\right)  \left (
\frac{\partial ^2 \xi^{\mu_1}}{\partial x^{\alpha}
\partial x^{\beta}}
+ \frac{\partial ^2 \xi^{\mu_1}}{\partial x^{\beta}
\partial x^{\alpha}}
\right)  - ( \beta \leftrightarrow \gamma)\\
&&\\
&&
+ \frac{1}{4}
\frac{\partial x^{\delta}}{\partial \xi^{\kappa}}
\frac{\partial \xi^{\mu}}{\partial x^{\alpha}}
g^{\alpha_1 \varepsilon} \eta_{\mu \nu}
\left (
\frac{\partial ^2 \xi^{\nu}}{\partial x^{\beta}
\partial x^{\varepsilon}}
-\frac{\partial ^2 \xi^{\nu}}{\partial x^{\varepsilon}
\partial x^{\beta}}
\right)  \left (
\frac{\partial ^2 \xi^{\kappa}}{\partial x^{\gamma}
\partial x^{\alpha_1}}
+ \frac{\partial ^2 \xi^{\kappa}}{\partial x^{\alpha_1}
\partial x^{\gamma}}
\right)  - ( \beta \leftrightarrow \gamma)\\
&&\\
&&
+ \frac{1}{4}
\frac{\partial x^{\delta}}{\partial \xi^{\kappa}}
\frac{\partial \xi^{\nu}}{\partial x^{\beta}}
g^{\alpha_1 \varepsilon} \eta_{\mu \nu}
\left (
\frac{\partial ^2 \xi^{\mu}}{\partial x^{\alpha}
\partial x^{\varepsilon}}
-\frac{\partial ^2 \xi^{\mu}}{\partial x^{\varepsilon}
\partial x^{\alpha}}
\right)  \left (
\frac{\partial ^2 \xi^{\kappa}}{\partial x^{\gamma}
\partial x^{\alpha_1}}
+ \frac{\partial ^2 \xi^{\kappa}}{\partial x^{\alpha_1}
\partial x^{\gamma}}
\right)  - ( \beta \leftrightarrow \gamma)\\
&&\\
&&
+ \frac{1}{4}
g^{\delta \varepsilon} \eta_{\lambda \mu_1}
\left (
\frac{\partial ^2 \xi^{\lambda}}{\partial x^{\gamma}
\partial x^{\varepsilon}}
-\frac{\partial ^2 \xi^{\lambda}}{\partial x^{\varepsilon}
\partial x^{\gamma}}
\right)  \left (
\frac{\partial ^2 \xi^{\mu_1}}{\partial x^{\alpha}
\partial x^{\beta}}
+ \frac{\partial ^2 \xi^{\mu_1}}{\partial x^{\beta}
\partial x^{\alpha}}
\right)  - ( \beta \leftrightarrow \gamma) ,
\ea
\ee

\be \label{5.26}
\ba{rcl}
G^{\delta}_{(8)\alpha\beta\gamma}
& = &
\frac{1}{4}
\frac{\partial \xi^{\mu}}{\partial x^{\alpha}}
\frac{\partial \xi^{\lambda}}{\partial x^{\gamma}}
g^{\alpha_1 \varepsilon } g^{\delta \varepsilon_1}
\eta_{\mu \nu} \eta_{\lambda \mu_1}
\left (
\frac{\partial ^2 \xi^{\nu}}{\partial x^{\beta}
\partial x^{\varepsilon}}
-\frac{\partial ^2 \xi^{\nu}}{\partial x^{\varepsilon}
\partial x^{\beta}}
\right)  \left (
\frac{\partial ^2 \xi^{\mu_1}}{\partial x^{\alpha_1}
\partial x^{\varepsilon_1}}
- \frac{\partial ^2 \xi^{\mu_1}}{\partial x^{\varepsilon_1}
\partial x^{\alpha_1}}
\right)  - ( \beta \leftrightarrow \gamma)\\
&&\\
&& + \frac{1}{4}
\frac{\partial \xi^{\mu}}{\partial x^{\alpha}}
\frac{\partial \xi^{\mu_1}}{\partial x^{\alpha_1}}
g^{\alpha_1 \varepsilon_1 } g^{\delta \varepsilon}
\eta_{\mu \nu} \eta_{\lambda \mu_1}
\left (
\frac{\partial ^2 \xi^{\nu}}{\partial x^{\beta}
\partial x^{\varepsilon_1}}
-\frac{\partial ^2 \xi^{\nu}}{\partial x^{\varepsilon_1}
\partial x^{\beta}}
\right)  \left (
\frac{\partial ^2 \xi^{\lambda}}{\partial x^{\gamma}
\partial x^{\varepsilon}}
- \frac{\partial ^2 \xi^{\lambda}}{\partial x^{\varepsilon}
\partial x^{\gamma}}
\right)  - ( \beta \leftrightarrow \gamma)\\
&&\\
&& + \frac{1}{4}
\frac{\partial \xi^{\nu}}{\partial x^{\beta}}
\frac{\partial \xi^{\lambda}}{\partial x^{\gamma}}
g^{\alpha_1 \varepsilon} g^{\delta \varepsilon_1}
\eta_{\mu \nu} \eta_{\lambda \mu_1}
\left (
\frac{\partial ^2 \xi^{\mu}}{\partial x^{\alpha}
\partial x^{\varepsilon}}
-\frac{\partial ^2 \xi^{\mu}}{\partial x^{\varepsilon}
\partial x^{\alpha}}
\right)  \left (
\frac{\partial ^2 \xi^{\mu_1}}{\partial x^{\alpha_1}
\partial x^{\varepsilon_1}}
- \frac{\partial ^2 \xi^{\mu_1}}{\partial x^{\varepsilon_1}
\partial x^{\alpha_1}}
\right)  - ( \beta \leftrightarrow \gamma)\\
&&\\
&&+ \frac{1}{4}
\frac{\partial \xi^{\nu}}{\partial x^{\beta}}
\frac{\partial \xi^{\mu_1}}{\partial x^{\alpha_1}}
g^{\alpha_1 \varepsilon } g^{\delta \varepsilon_1}
\eta_{\mu \nu} \eta_{\lambda \mu_1}
\left (
\frac{\partial ^2 \xi^{\mu}}{\partial x^{\alpha}
\partial x^{\varepsilon}}
-\frac{\partial ^2 \xi^{\mu}}{\partial x^{\varepsilon}
\partial x^{\alpha}}
\right)  \left (
\frac{\partial ^2 \xi^{\lambda}}{\partial x^{\gamma}
\partial x^{\varepsilon_1}}
- \frac{\partial ^2 \xi^{\lambda}}{\partial x^{\varepsilon_1}
\partial x^{\gamma}}
\right)  - ( \beta \leftrightarrow \gamma).
\ea
\ee
In the above relations, symbol $( \beta \leftrightarrow \gamma)$
represents the term with exchange two index $\beta$ and $\gamma$
of the  preceding term. From above expressions of
$G^{\delta}_{\alpha\beta\gamma}$, we find that if
eqs.(\ref{5.6} -- \ref{5.7}) hold, $G^{\delta}_{\alpha\beta\gamma}$
strictly vanishes and the transformation of curvature tensor is
covariant; i.e., the curvature tensor transforms covariantly
under the first kind of general coordinate transformations. But under
the second kind of general
coordinate transformations, the curvature tensor
does not transform covariantly. It means that, under the second
kind of general coordinate transformations,
curvature tensor is not a real tensor. Correspondingly,
the transformation  rule of Recci tensor is
\be \label{5.27}
R_{\mu\nu} \to
{\mathbb R}_{\alpha\beta}
=\frac{\partial \xi^{\mu}}{\partial x^{\alpha}}
\frac{\partial \xi^{\nu}}{\partial x^{\beta}}
R_{\mu\nu}
+ G_{\alpha\beta},
\ee
where
\be \label{5.28}
G_{\alpha \beta} = G^{\gamma}_{\alpha\gamma\beta}.
\ee
Similarly, Recci tensor is not a real tensor under the second kind
of general  coordinate transformations.
The transformation of curvature scalar $R$ is
\be \label{5.29}
R \to {\mathbb R} = R + G,
\ee
where
\be \label{5.30}
G = g^{\alpha\beta} G_{\alpha\beta}.
\ee
Under the second kind of general coordinate
transformations, $G$ does not vanish,
so the curvature scalar $R$ change its magnitude, which means that
the space-time structure was changed.
The second kind of general coordinate
transformations can transform a flat space-time into a curved
space-time or vice versa. So, a transformation which changes
curvature scalar must belong to the
second kind of general coordinate
transformations, for example, the transformation between local
inertial reference system and a curved space-time is a second kind
of general coordinate transformations.
\\

Covariant derivatives are defined by eqs.(\ref{3.41} -- \ref{3.42}).
Under transformation eq.(\ref{5.2}), the transformation rules for
covariant derivatives are given by eqs.(\ref{3.43} -- \ref{3.44}).
For the second kind of general
coordinate transformations, these covariant
derivatives are not real covariant.
Under transformation eq.(\ref{5.2}), the transformation rule for
continuity equation is given by eq.(\ref{4.1}), that for geodesic equation
is given by eq.(\ref{4.3}), that for parallel transport equation
is given by eq.(\ref{4.8}), that for energy-momentum conservation
equation is given by eq.(\ref{4.20}) and that for Einstein  field
equation is given by eq.(\ref{4.25}). We could see that all these
basic physical equations can not preserve their forms under
the second kind of general coordinate transformations. The violation of general
covariance originates from the violation of eqs.(\ref{5.6} -- \ref{5.7}),
which cause $B^{\gamma}_{\alpha\beta}$ and $G^{\delta}_{\alpha\beta\gamma}$
not vanish. \\

\section{Problems on the Foundations of General Relativity}
\setcounter{equation}{0}

The violation of general covariance causes serious problems to the
foundations of general relativity. It is known that the {\it a priori} foundations
of general relativity are the equivalence principle and the principle
of general covariance, and these principles are directly related
to the property of general covariance under general coordinate
transformations. In general relativity, these
two principles tell us how to obtain the laws of nature in an
arbitrary curved space-time. The equivalence
principle tells us that, in any
point of an arbitrary curved space-time, there must exist a local
inertial coordinate system and the laws of nature in it take
the same form as that in unaccelerated  Cartesian coordinate
system in absence of gravity. Then the principle of general covariance
tells us that the laws of nature in an arbitrary curved space-time
are obtained through a general coordinate transformation from the
local inertial coordinate system to the curved space-time. But
the coordinate transformation from a local inertial coordinate
system to any curved space-time belongs to the second kind of general
coordinate transformations. Under the second kind of general coordinate
transformations, all basic physical equations are not covariant.
All physical equations given by this coordinate transformation are
different from those in general relativity. So, there is a
fundamental problem: what are the {\it a priori} foundations of
general relativity? All basic physical equations
in general relativity can not be obtained based on the equivalence
principle and the principle of general covariance, for equations
deduced based on these two principles are different from those
in general relativity. \\

According to the theory of special relativity, the laws of nature
in an inertial reference system are exactly known. If the equivalence
principle strictly holds, the laws of nature in a local inertial
coordinate system are also exactly known. If the nature of gravity
is space-time geometry, the laws of nature in different kinds of
curved space-time must be related to each other through a
general coordinate transformation from one curved
space-time to another curved space-time. Therefore, the laws of
nature in a curved space-time can be obtained through a
coordinate transformation from a local inertial coordinate
system to the curved space-time. If this procedure is true,
we will encounter two essential problems: (1) the basic physical
equations obtained in this way are different from those in general
relativity; (2) the basic physical equations
can not be expressed in terms of physical observable,
metric tensor and their space-time derivatives. In other words,
if basic physical equations are determined in this way, an
observer in a curved space-time can not determined the exact
forms of these basic physical equations, for he has no information
on the transformation matrix
$\frac{\partial \xi^{\mu}}{\partial x^{\alpha}}$
and $\frac{\partial x^{\alpha}}{\partial \xi^{\mu}}$. Furthermore,
the transformation matrix from a local inertial coordinate
system to a curved space-time is not unique\cite{wu8}, all
physical equations in the new coordinate system are also not
unique, which is another serious problem on geometry picture
of gravity. \\

Because basic physical equations are not covariant under a
general coordinate transformation, what is the symmetry of
gravity in general relativity?\\

In a word, because of the violation of general covariance
under the second kind of general coordinate transformations, both
the equivalence principle and the principle of general covariance
can not be regarded as {\it a priori} foundations of general relativity. \\

\section{Discussions}
\setcounter{equation}{0}

In this paper, violation of general covariance under the second
kind of general coordinate transformations
is studied and problems on foundations
of general relativity is discussed. If gravity is space-time geometry,
basic physical equations in curved space-time of different kind should
be related each other through a general coordinate transformation, for
these curved space-time can be related each other through a general
coordinate transformation. But the basic physical equations are not covariant
under a general coordinate transformation, all basic physical equations
obtained from the equivalence principle and the principle of general
covariance are different from those in general relativity.
Besides, the forms of basic physical equations
can not be uniquely determined through general coordinate transformations,
for the transformation matrix from a local inertial coordinate
system to a curved space-time is not unique. So, both the equivalence
principle and the principle of general covariance are not {\it a priori}
foundations of general relativity. The violation of general
covariance causes serious problems to the foundations of general
relativity. \\

Another essential problem related to general relativity is that
quantum general relativity is not perturbatively renormalizable.
At present, we can not set up a self-consistent quantum
theory of general relativity.
\\

Because of the great achievement of QCD, QED and unified electroweak
theory in particle physics, it is generally believed that the
common nature of all fundamental interactions in nature is gauge
theory. Based on this belief, quantum gauge theory of gravity
is proposed which is renormalizable in 4-dimensional Minkowski
space-time\cite{wu1,wu2,wu3,wu4}. Quantum gauge theory of gravity
is set up in the physics picture of gravity and gravity is treated
as a kind of physical interactions, not space-time geometry, so it
has a clear physical picture. In quantum gauge theory of gravity,
four different kind of fundamental interactions in nature can
be unified in a simple and beautiful way\cite{wu5,wu6,wu7}. All
these are advantage of quantum gauge theory of gravity.  \\

Furthermore, quantum gauge theory of gravity does not have the
above fundamental problems. In quantum gauge theory of gravity,
gravity is treated as a kind of fundamental interactions, not
space-time geometry. Physics is formulated in flat 4-dimensional
Minkowski space-time, not in arbitrary curved space-time. So,
in quantum gauge theory of gravity, we do not encounter the
problems caused by the second kind of general
coordinate transformations.
Basic physical equations in quantum gauge
theory of gravity are obtained through gauge principle and
action principle, not through a general coordinate transformation.
This is another  advantage of quantum gauge theory
of gravity.  \\

Different gravity theory has different kind of symmetry, such as
general relativity has the symmetry of general covariance,
canonical quantum gravity has local Lorentz symmetry and quantum
gauge theory of gravity has gravitational gauge symmetry.
So, what is the true symmetry of gravitational interactions?
According to discussions in this paper and literature \cite{wu8},
basic physical equations can not preserve their forms under
either general coordinate transformations or local Lorentz
transformations. So, the symmetry of gravitational interactions
are not the symmetry of general covariance, nor the local Lorentz
symmetry. Therefore, the correct symmetry of gravitational
interactions should be gravitational gauge symmetry. On the
other hand, gravitational gauge symmetry is a natural
requirement of gauge principle\cite{wu1,wu2,wu3,wu4}. \\

\lbrack Acknowledgement \rbrack
The author would like thank Prof. Dahua Zhang,
Prof. Hanqing Zheng and Prof. Zhipeng Zheng for useful
discussions. \\

\end{document}